\documentclass[a4paper,11pt]{article}
\pdfoutput=1 

\usepackage{jcappub} 
\usepackage[T1]{fontenc} 

\title{\boldmath Decoherence-effects in the neutrino-mixing mechanism: active and sterile neutrinos in the three flavor scheme} 

\author[a,b]{M. E. Mosquera,}
\author[b,1]{O. Civitarese\note{Corresponding author.}}

\affiliation[a]{Facultad de Ciencias Astron\'{o}micas y Geof\'{\i}sicas, Universidad Nacional de La Plata, \\ Paseo del Bosque, (1900) La Plata, Argentina}
\affiliation[b]{Department of Physics, University of La Plata, \\ c.c. 67 (1900), La Plata, Argentina}

\emailAdd{mmosquera@fcaglp.unlp.edu.ar}
\emailAdd{osvaldo.civitarese@fisica.unlp.edu.ar}

\keywords{decoherence, neutrino fluxes}

\abstract{In this work we analyse the effects of quantum decoherence upon the precession of the neutrino polarization vector in a supernovae-like environment. In order to perform the study we have determined the time-dependence of the polarization vector by solving the equation of motion for different neutrino-mixing schemes. The results of the calculations show that the onset of decoherence depends strongly on the parameters of the adopted mixing scheme. As examples we have considered : a) the mixing between active neutrinos and b) the mixing between active end sterile neutrinos.}

\begin{document}
\maketitle

\section{Introduction}
\label{Intro}

The study of neutrinos produced in scenarios of astrophysical interest is a cross disciplinary subject which relates aspects of cosmology, particle and 
nuclear physics. Neutrinos produced in different processes such as relic neutrinos, supernova neutrinos, neutrinos produced 
in micro-quasar's jets, among others, might arrive to Earth without being affected by local interactions. They 
exhibite oscillations in flavor, a phenomena predicted long ago by  Pontecorvo. The review of Ref.\cite{bilenky} gives a nice and rigorous introduction to the subject. The 
parameters which are associated to neutrino oscillations have been determined experimentally  \cite{forero14,gonzalez14,capozzi14,capozzi16,qian15}, thought 
the absolute scale of the neutrino mass is still unknown. This value can eventually be determined if the neutrinoless double beta decay is observed \cite{0nubb}. 

From quantum mechanics one can determine the evolution of pure states 
into mixed states due to interactions with a background \cite{schlosshauer04,zurek81}. Therefore, and 
due to the quantum nature of neutrinos, it is expected that decoherence can modify the neutrino's wave packet in their way from the source to the detectors 
\cite{schlosshauer04,zurek81,zurek82,raffelt10,smirnov}. 

In this work we focus our attention in the study of the effects due to decoherence acting on neutrinos produced in supernovae, for different 
neutrino-mixing schemes. In Ref. \cite{mosquera18} we have studied 
the onset of decoherence on the spectrum of neutrinos produced in micro-quasar jets. 
The calculation was performed using two active neutrinos and in the one-active plus one-sterile neutrino ($1+1$)-scheme. In this work we have extended the formalism of Ref. \cite{raffelt10} in order to include three neutrino states. As initial distribution function we have used a Fermi-Dirac with values of temperature and mean energy adequate to the supernova environment \cite{balantekin04,tamborra12}. We have considered free as well as interacting neutrinos.

The work is organized as follows. In Section \ref{deco} we briefly introduce the formalism needed to calculate the time evolution of the energy spectrum for the superposition of three neutrino-mass eigenstates. In Section \ref{espectra} we describe the interactions  and initial conditions used to calculate the time evolution of the polarization vector in presence of decoherence in the three active neutrinos scheme and $2+1$-scheme (two active plus one sterile neutrinos). The results of the calculations are presented and discussed in Section \ref{resultados}. Our conclusions are drawn in Section \ref{conclusion}.

\section{Formalism}
\label{deco}

The time evolution of the occupation number of neutrinos is governed by the equation of motion \cite{raffelt10}
\begin{eqnarray}
\label{eq}
\imath \dot{\rho_f}&=&\left[\frac{M^2 c^4}{2E \hbar}+ \frac{\sqrt{2}G_F \left(\hbar c\right)^3 }{\hbar} \,\rho, \, \rho_f \,\, \right] \, \, ,
\end{eqnarray}
where the squared brackets reads for the commutator, the dot represents the time derivative and $M^2 c^4$ is the mass-squared matrix in the flavor basis. The density matrix in the flavor basis is $\rho_f$, and $\rho$ is a matrix whose diagonal terms are the neutrino number densities, $E$ stands for the neutrino energy and $G_F$ is the Fermi constant. 

Given an unitary mixing matrix $U$, which relates the mass-eigenstates with the flavor-eigenstates, one can write the mass-squared matrix in the flavor basis as
\begin{eqnarray}
M^2 c^4&=&m_1^2 c^4 I_3 + \left(m_3^2-m_1^2\right)c^4 \left(
\begin{array}{ccc}
\Delta U_{12}^2+U_{13}^2 & \Delta U_{12} U_{22}+U_{13}U_{23} & \Delta U_{12}U_{32}+U_{13} U_{33} \\
\Delta U_{12} U_{22}+U_{13}U_{23} &\Delta U_{22}^2+U_{23}^2 &\Delta U_{22}U_{32}+U_{23} U_{33} \\
\Delta U_{12}U_{32}+U_{13} U_{33}  &\Delta U_{22}U_{32}+U_{23} U_{33} &\Delta U_{32}^2+U_{33}^2
\end{array}
\right) \, \, \, , \nonumber
\end{eqnarray}
where $I_3$ is the $3\times3$ identity matrix, $m_i$ stands for the mass of the $i$-eigenstate and $\Delta=\frac{m_2^2-m_1^2}{m_3^2-m_1^2}$. This matrix can be written in terms of the Gell-Mann matrices $\lambda_i$ as
\begin{eqnarray}
M^2 c^4&=&\frac{1}{3} {\rm Tr}\left(M^2 c^4\right) I_3 + \frac{\left(m_3^2-m_1^2\right)c^4}{2} {\bar{B}} \cdot {\bar{\lambda}} \, \, \, .
\end{eqnarray}
By comparing the two previous equations one obtains the expression of the vector $\bar{B}$, which in matrix form
reads
\begin{eqnarray}
\label{b}
\bar{B}&=&\left(
\begin{array}{c}
2\left(\Delta U_{12} U_{22}+U_{13}U_{23}\right)\\
0\\
\Delta \left(U_{12}^2-U_{22}^2\right)+U_{13}^2-U_{23}^2\\
2\left(\Delta U_{12} U_{32}+U_{13}U_{33}\right)\\
0\\
2\left(\Delta U_{22} U_{32}+U_{23}U_{33}\right)\\
0\\
\sqrt{3}\left(\Delta \left(\frac{1}{3}- U_{32}^2\right)+\frac{1}{3}- U_{33}^2\right)
\end{array}
\right) \, \, \, .
\end{eqnarray}
 
The matrix $\rho_f$ can be expressed in terms of Gell-Mann matrices $\lambda$ as:
\begin{eqnarray}
\rho_f&=&\frac{1}{3} {\rm Tr \left(\rho_f\right)} I_3 + \frac{1}{2} {\bar{P}_w} \cdot {\bar{\lambda}} \, \, \, .
\end{eqnarray}
In the previous expression $\bar{P}_w $ is the polarization vector in the flavor basis. With the use of  Eq. (\ref{eq}) it is
 transformed into
\begin{eqnarray}\label{pol}
\dot{\bar{P}}_w&=&\left(w \bar{B} + \chi \, \bar{P} \right) \times \bar{P}_w \, \, ,
\end{eqnarray}
where $w=\frac{\Delta m_{31}^2c^4}{2E\hbar}$, and  $\Delta m_{31}^2c^4=m_3^2 c^4-m_1^2 c^4$ is the mass-squared difference between  mass eigenstates. 
The vector $\bar{P}=\int \bar{P}_w\, {\rm d}w$ is the total (or global) polarization vector. The cross product is the inner product between 
two vectors with the structure constant of the SU(3) group. The vector $\bar{B}$ fixes a certain orientation of the background (Eq. (\ref{b})). 
The parameter $\chi$ stands for the strength of the neutrino-neutrino interaction \cite{raffelt10}. 
The initial condition for $\bar{P}_w$, is given by
\begin{eqnarray}
\label{pw0}
\bar{P}_w(0)&=&\left(
\begin{array}{c}
0\\
0\\
g_{\nu_e}(w)-g_{\nu_\mu}(w) \\
0\\
0\\
0\\
\left(g_{\nu_e}(w)+g_{\nu_\mu}(w)-g_{\nu_\tau}(w)\right)/\sqrt{3}
\end{array}
\right) \,\, \, ,
\end{eqnarray}
where $g_{\nu_e}$, $g_{\nu_\mu}$ and $g_{\nu_\tau}$ are the electron, 
muon and tau-neutrino spectral normalized functions. These distribution functions are defined in Section \ref{espectra}.

The quantity which measures decoherence is the order parameter $R(t)$ which is defined as the length of the component
 of $\bar{P}$ perpendicular $\left(\bar{P}_\perp (t)\right)$ to the vector $\bar{B}$  
 normalized at $t=0$. In this way
\begin{eqnarray}
\label{rtheta}
R(t)&=&\frac{\left|\bar{P}_\perp (t)\right|}{\left|\bar{P}_\perp (0)\right|} \, \, \, .
\end{eqnarray}

For the case of three active neutrinos the mixing matrix reads
\begin{eqnarray}
U&=&\left(
\begin{array}{ccc}
c_{12} c_{13} & s_{12} c_{13} & s_{13} \\
-s_{12} c_{23}- c_{12} s_{23} s_{13} & c_{12} c_{23} -s_{12} s_{23} s_{13} & c_{13} s_{23}\\
s_{12} s_{23}-c_{12} c_{23} s_{13} & -c_{12} s_{23} -s_{12} c_{23} s_{13} & c_{13} c_{23}
\end{array}
\right) \, \, \, ,
\end{eqnarray}
where $c_{ij}$ and $s_{ij}$ stand for $\cos \theta_{ij}$ and $\sin \theta_{ij}$ respectively. For the $2+1$-scheme we have used
\begin{eqnarray}
U&=&\left(
\begin{array}{ccc}
c_{12} c_{14} &s_{12} &c_{12} s_{14}\\
-s_{12} c_{14} & c_{12} & -s_{12} s_{14}\\
-s_{14} & 0 & c_{14}
\end{array}
\right) \, \, \, ,
\end{eqnarray}
and, in this case, the frequency is defined as $w=\frac{\Delta m_{41}^2c^4}{2E\hbar}$. 

\section{Neutrino spectra}
\label{espectra}

For the calculation of the factor of decoherence we have used Gaussian-like functions
to model the interactions between neutrinos. We assume that the interactions depend on the frequency of the interacting neutrinos
\cite{raffelt10,mosquera18}.

\subsection{Gaussian spectrum}
\label{gauss}

The spectra of the electron and muon-type neutrinos are given by the functions 
\begin{eqnarray}
g_{\nu_e}(w)&=& \frac{1}{2\sqrt{2\pi}}e^{-(w-5)^2/2} + \frac{1}{2\sqrt{2\pi}}e^{-(w+5)^2/2}\, \, \, , \nonumber \\
g_{\nu_\mu}(w)&=& \frac{1}{2\sqrt{2\pi}}e^{-(w-6)^2/2} + \frac{1}{2\sqrt{2\pi}}e^{-(w+6)^2/2}\, \, \, ,\nonumber \\
g_{\nu_\tau}(w)&=& 0\, \, \,,
\end{eqnarray}
which are non-overlapping functions of the frequency.

\subsection{Supernova neutrino spectrum}
\label{sn}

Neutrinos produced in a supernova core-bounce obey a Fermi-Dirac distribution function 
\begin{eqnarray}
\rho_{\nu_\alpha}(E) dE&=& \frac{1}{ \pi^2\left(\hbar c\right)^3}\frac{dE\;\;E^2}{1+e^{E/T_{\nu_\alpha}}}\, \, \, ,
\end{eqnarray}
where the temperature $T_{\nu_\alpha}$ is related to the mean energy of the $\alpha$-neutrino specie $\left(\nu_\alpha\right)$ $T_{\nu_\alpha}=\left< E_{\nu_\alpha}\right>/3.151$. The values of the mean energies were extracted from Ref. \cite{balantekin04,tamborra12}, that is $\left< E_{\nu_e}\right>=10 \, {\rm MeV}$,  $\left< E_{\overline{\nu}_e} \right>=15 \, {\rm MeV}$ and $\left< E_{\nu_x}\right>=\left< E_{\overline{\nu}_x}\right>=24 \, {\rm MeV}$, where the sub-index $x$ represents the muon or tau-neutrino. 

The distribution functions of  neutrinos of a given flavor, as function of the frequency, are written:
\begin{eqnarray}\label{distnu}
g_{\nu_e}(w)&=& \frac{1}{ \pi^2\left(\hbar c\right)^3} \left(\frac{\Delta m_{31}^2 c^4}{2\hbar}\right)^3 \frac{1}{w^4} \frac{1}{1+e^{\Delta m_{31}^2 c^4/(2\hbar T_{{\nu_e},{\overline{\nu}_e}} |w|)}}\, \, \, , \nonumber \\
g_{\nu_\mu}(w)&=& \frac{1}{ \pi^2\left(\hbar c\right)^3} \left(\frac{\Delta m_{31}^2 c^4}{2\hbar}\right)^3 \frac{1}{w^4} \frac{1}{1+e^{\Delta m_{31}^2 c^4/(2\hbar T_{{\nu_\mu},{\overline{\nu}_\mu}} |w|)}}\, \, \, ,\nonumber \\
g_{\nu_\tau}(w)&=& \frac{1}{ \pi^2\left(\hbar c\right)^3} \left(\frac{\Delta m_{31}^2 c^4}{2\hbar}\right)^3 \frac{1}{w^4} \frac{1}{1+e^{\Delta m_{31}^2 c^4/(2\hbar T_{{\nu_\tau},{\overline{\nu}_\tau}} |w|)}}\, \, \, .
\end{eqnarray}
These functions are normalized to the total neutrino density
\begin{eqnarray}\label{nu}
n_\nu&=& \int_{-\infty}^\infty \left(g_{\nu_e}(w)+g_{\nu_\mu}(w)+g_{\nu_\tau}(w)\right) \, {\rm d}w\, \, \, .
\end{eqnarray}
The positive (negative) values of $w$ in the integral of Eq.(\ref{nu}) give the contribution of neutrinos (anti-neutrinos) to the neutrino density. For the calculations in the $2+1$-scheme we have assumed that initially there is not sterile neutrinos in the supernova environment, and  in Eqs.(\ref{distnu}) the factor $\Delta m_{31}^2c^4$ is replaced by $\Delta m_{41}^2 c^4$.

\section{Results}
\label{resultados}

To compute the order parameter $R$, we have considered two different scenarios: i) three active neutrinos and ii) two active neutrinos and one sterile neutrino, as said before. The neutrino mixing parameters were extracted from Ref. \cite{mixing}
\begin{eqnarray}
\Delta m_{21}^2c^4&=&7.53 \times 10^{-5} \, {\rm eV}^2 \, \, \, , \nonumber \\
\Delta m_{31}^2c^4&=&2.45 \times 10^{-3} \, {\rm eV}^2 \, \, \, , \nonumber \\
\sin^2 \theta_{12}&=&0.307 \, \, \, , \nonumber \\
\sin^2 \theta_{23}&=&0.510 \, \, \, , \nonumber \\
\sin^2 \theta_{13}&=&0.021 \, \, \, .
\end{eqnarray}
For the $2+1$-scheme scenario we have used different values of the mass square difference and the mixing angle. As initial condition, for this scheme, we assume that sterile neutrinos are not produced in the supernova, as said before. 

\subsection{Gaussian spectrum}

\begin{figure}[!h]
\begin{center}
\includegraphics[width=200 pt,angle=-90]{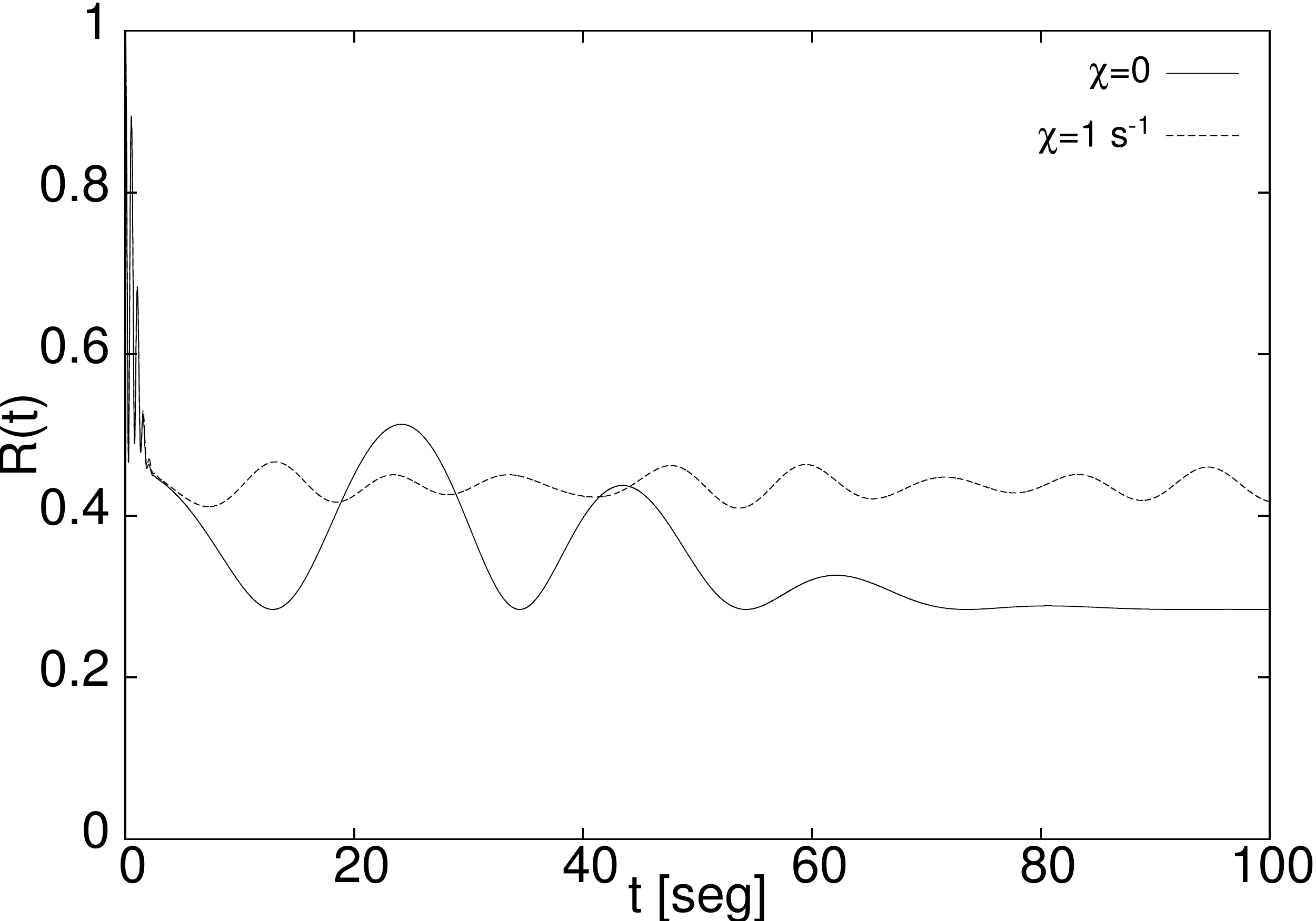}
\end{center}
\caption{Order parameter for Gaussian-like distribution functions, as a function of time for two different values of the neutrino interaction $\chi$ considering three active neutrinos. Solid line: $\chi=0$, dashed line: $\chi= 1 \, {\rm s}^{-1}$.}
\label{gauss-activo}
\end{figure}

We have calculated the order parameter for the Gaussian-like distribution functions presented in Section \ref{gauss}. In Fig. \ref{gauss-activo} we show the results for the parameter $R$ as a function of time using two different values of the neutrino interaction $\chi$. If the neutrino interaction is turned off the order parameter acquires a non-zero, almost constant value value of $\left(R\sim 0.28\right)$. If one activates the neutrino-neutrino interaction the decoherence reduces, that is $R$ acquires a higher value at large time ($\left(R\sim 0.4\right)$. Thus, the length of the vector $\bar{P_\perp}$ increases with the strength of the neutrino-neutrino interaction.

\begin{figure}[!h]
\begin{center}
\includegraphics[width=190 pt,angle=-90]{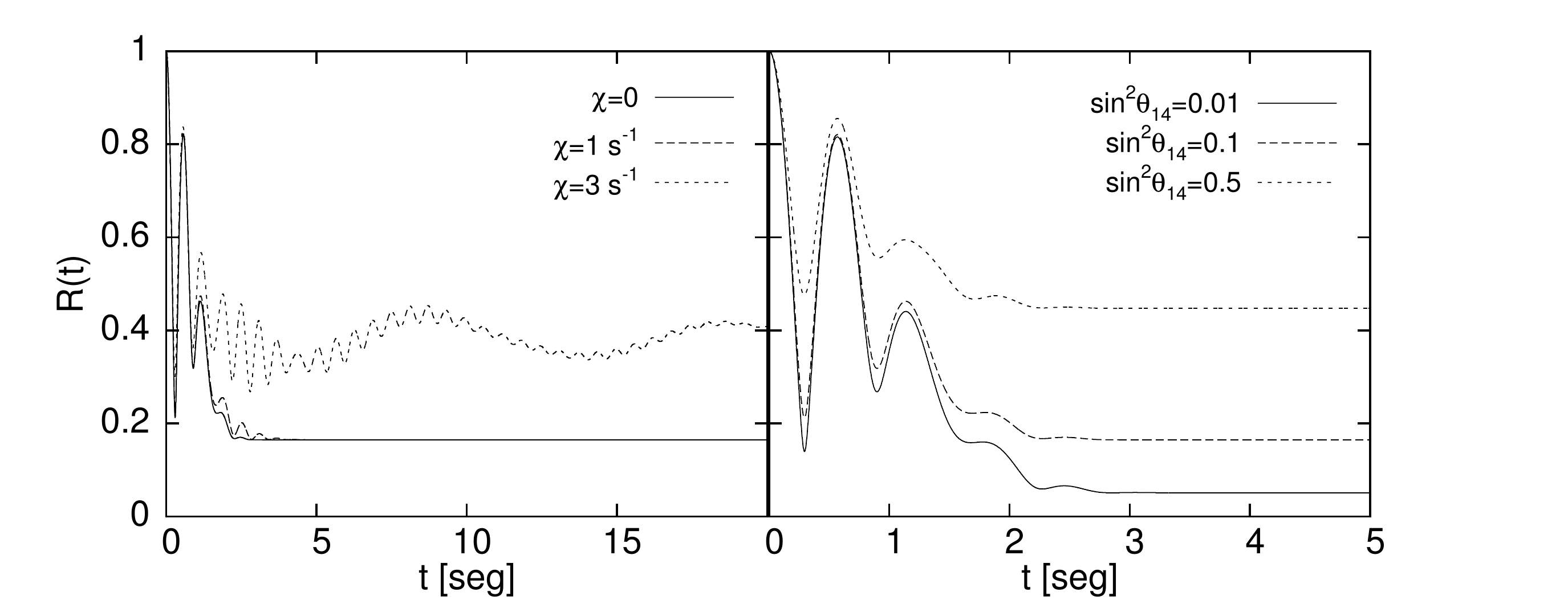}
\end{center}
\caption{Order parameter factor as a function of time for Gaussian-like distribution functions in the $2+1$-scheme. The square mass difference for both figures is $\Delta m_{41}^2 c^4= 1 \, {\rm eV}^2$. Left figure: the mixing angle was fixed at $\sin^2 \theta_{14}=0.1$ and the value of $\chi$ is variable (solid line: $\chi=0$; dashed line: $\chi=1 \, {\rm s}^{-1}$; dotted line: $\chi=3 \, {\rm s}^{-1}$). Right figure: the neutrino interaction is turned off ($\chi=0$), and the value of the mixing angle is variable (solid line: $\sin^2 \theta_{14}=0.01$; dashed line: $\sin^2 \theta_{14}=0.1$; dotted line: $\sin^2 \theta_{14}=0.5$).}
\label{gauss-esteril}
\end{figure}

The results of the inclusion of a sterile neutrino in the computation of the order parameter ($2+1$-scheme) are shown in Figure \ref{gauss-esteril}. In the left inset of Figure \ref{gauss-esteril} we present the results for $\Delta m_{41}^2 c^4= 1 \, {\rm eV}^2$ and $\sin^2 \theta_{14}=0.1$ for different values of the interaction $\chi$. Once again, the larger the value of $\chi$, the larger the value of the order parameter. In the right inset of Fig. \ref{gauss-esteril} we show the effect of the variation of the mixing angle upon the length of the vector $\bar{P_\perp}$. In all the cases, the order parameter decreases its value with time, however the final value is strongly dependent on the active-sterile neutrino mixing angle. 

\subsection{Supernova neutrino spectrum}

\begin{figure}[!h]
\begin{center}
\includegraphics[width=200 pt,angle=-90]{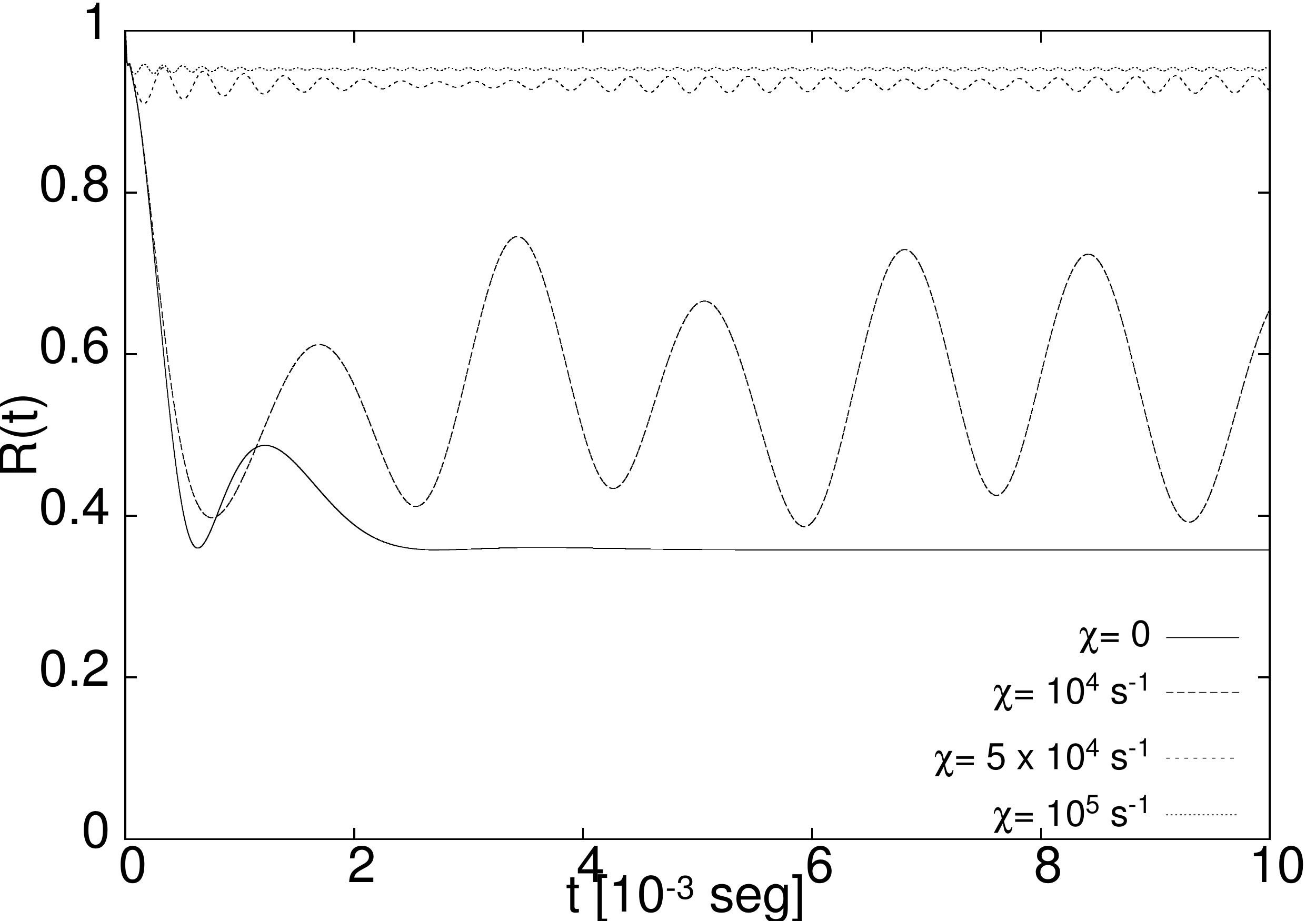}
\end{center}
\caption{Order parameter computed in the three active neutrino scheme, as a function of the time for different values of the neutrino interaction $\chi$. Solid line: $\chi=0$; long dashed line $\chi=10^4 \, {\rm s}^{-1}$; small dashed line $\chi= 5 \times 10^4 \, {\rm s}^{-1}$; dotted line $\chi=10^5 \, {\rm s}^{-1}$.}
\label{sn-activo}
\end{figure}

Here we present the results for the order parameter calculated by using the Fermi-Dirac neutrino spectral functions of Section \ref{sn}. We show the length of the vector $\bar{P_\perp}$ in Figure \ref{sn-activo} for the three active neutrino scheme. As one can see, the order parameter is reduced to $0.36$ if  neutrino-interactions are neglected. When the interaction is turn on, the  polarization vector oscillates towards an asymptotic non-zero value which increases with $\chi$ (the strength  of the neutrino-neutrino interactions). The values of $\chi$ given in the figure are scaled to the values of the corresponding $\omega$
(see Eq.(\ref{pol})).

\begin{figure}[!h]
\begin{center}
\includegraphics[width=190 pt,angle=-90]{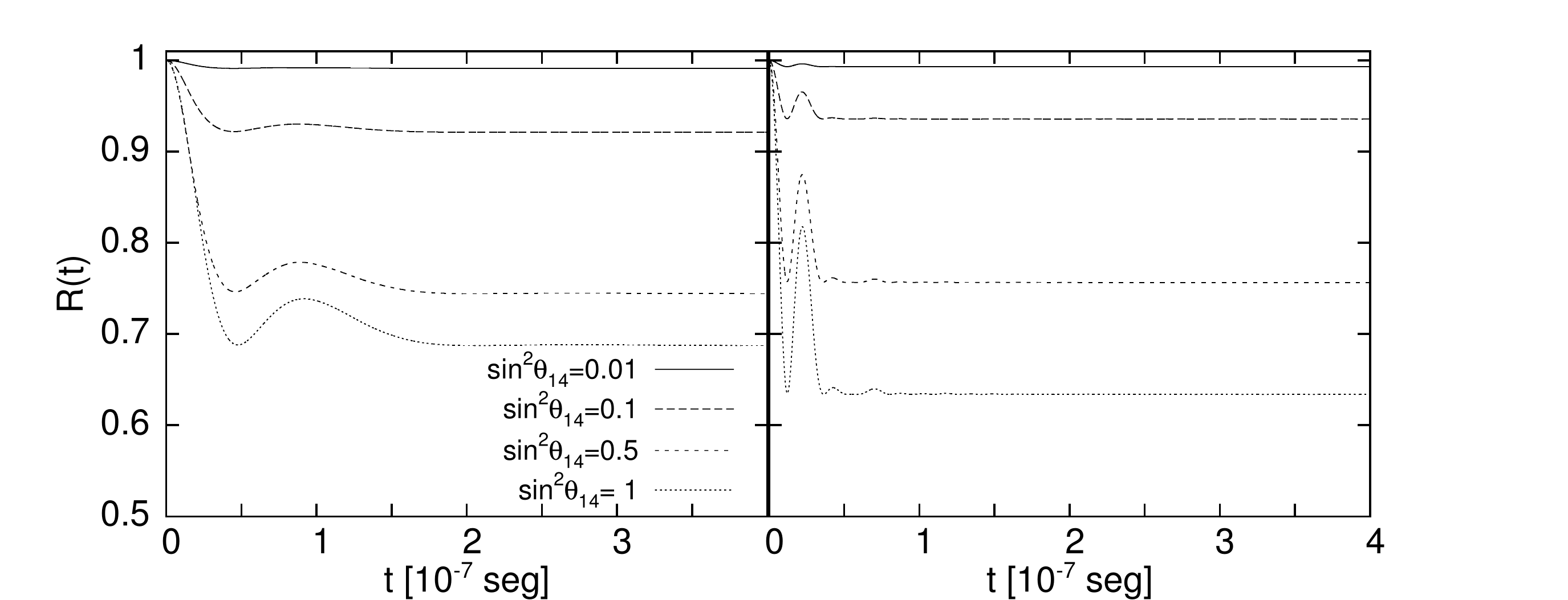}
\end{center}
\caption{$R(t)$ calculated in the $2+1$-scheme without neutrino-neutrino interactions, and for different values of the mixing angle. Left figure: $\Delta m_{41}^2 c^4= 1 \, {\rm eV}^2$; right figure: $\Delta m_{41}^2 c^4= 5 \, {\rm eV}^2$. Solid line: $\sin^2 \theta_{14}=0.01$; long dashed line: $\sin^2 \theta_{14}=0.1$; small dashed line: $\sin^2 \theta_{14}=0.5$; dotted line: $\sin^2 \theta_{14}=1$.}
\label{sn-esteril-dm}
\end{figure}

The results of the calculations of $R$ in the $2+1$-scheme for two different values of the mass square difference are shown in 
Figure \ref{sn-esteril-dm}. The left inset corresponds to $\Delta m_{41}^2 c^4= 1 \, {\rm eV}^2$, the right inset represents the results with $\Delta m_{41}^2 c^4= 5 \, {\rm eV}^2$. We have used different mixing angles and null neutrino-neutrino interactions in both calculations. If the mixing angle is $\pi/2$, the value of the order parameter is $0.69$ for the lower mass square difference, and $0.63$ for the larger value of $\Delta m_{41}^2 c^4$. If the mixing angle is smaller, the order parameter acquires a higher value.

\begin{figure}[!h]
\begin{center}
\includegraphics[width=190 pt,angle=-90]{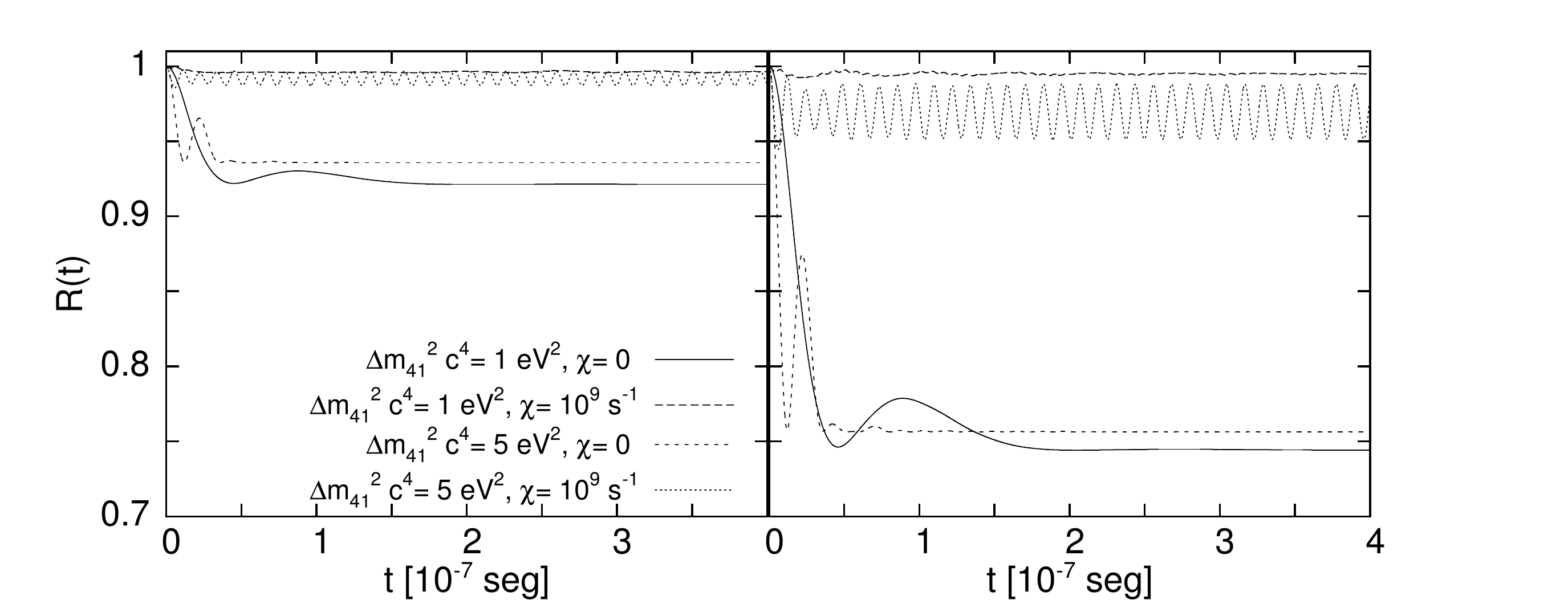}
\end{center}
\caption{Order parameter calculated in the $2+1$-scheme as a function of time for different values of the active-sterile neutrino oscillation parameters and neutrino-neutrino interaction. Left figure: $\sin^2 \theta_{14}= 0.5$; right figure: $\sin^2 \theta_{14}= 1$. Solid line: $\Delta m_{41}^2 c^4= 1 \, {\rm eV}^2$ and $\chi=0$; long dashed line: $\Delta m_{41}^2 c^4= 1 \, {\rm eV}^2$ and $\chi= 10^9 \, {\rm s}^{-1}$; small dashed line: $\Delta m_{41}^2 c^4= 5 \, {\rm eV}^2$ and $\chi=0$; dotted line: $\Delta m_{41}^2 c^4= 5 \, {\rm eV}^2$ and $\chi= 10^9 \, {\rm s}^{-1}$.}
\label{sn-esteril-sin}
\end{figure}

In Fig. \ref{sn-esteril-sin} we present the results of $R$ for two different mixing angles (left inset: $\sin^2 \theta_{14}= 0.5$, right inset: $\sin^2 \theta_{14}= 1$) in the $2+1$-scheme, for different values of the strength of the neutrino-neutrino interaction and $\Delta m_{41}^2 c^4$. When the neutrino interaction is turned off, the reduction of the length of the vector $\bar{P}_\perp$ is higher than the case with $\chi\neq 0$. 

\section{Conclusions}
\label{conclusion}

In this work we have studied the effects of decoherence  upon neutrino distribution functions for neutrinos produced in a supernova-like environment. This analysis was performed by extending the formalism presented in Ref. \cite{raffelt10}. Knowing the initial distribution function we have calculated the evolution of the polarization vector $\bar{P}$ and consequently of the order parameter $R$ as a function of time for different neutrino-schemes.

For the case of neutrino's Gaussian distribution functions, which model the density of neutrinos in the interior of the supernova, we have found that the polarization vector is reduced and oscillates around a non-zero value. The larger the value of the strength of the neutrino-neutrino interactions the larger is the asymptotic value of the polarization vector. Naturally, the asymptotic value of the polarization vector depends  on the mixing-angle between neutrino mass eigenstates.

This pattern appears also in the case of neutrinos produced in a supernova-like environment. If one sterile-neutrino specie is included in the calculations the effect is reduced and it becomes strongly dependent on the mixing angle and the squared mass difference. When a sterile neutrino is oscillating with a light active-neutrino the effects of the collective oscillations become noticeable at very small times, since the maximum value for the frequency $w$ is quite large due to the mass difference, as it is the realistic strength of neutrino-neutrino interactions.

The trend of the results suggests that the asymptotic value $R\approx 1$ may be reached when realistic values of strength of 
neutrino-neutrino interactions are used.

\section*{{\bf Acknowledgments}}
This work was supported by a grants (PIP-616) of the National Research Council of Argentina (CONICET), and by a research-grant of the National Agency for the Promotion of Science and Technology (ANPCYT) of Argentina. The authors are members of the Scientific Research Career of the CONICET.

\bibliography{bibliografia}
\bibliographystyle{jhep}
\end{document}